\documentclass{PoS}

\title{Star formation in the Local Group as seen by low-mass stars}

\ShortTitle{Star formation in the Local Group as seen by low-mass stars}

\author{\speaker{Guido De Marchi} \\
        European Space Research and Technology Centre, Keplerlaan 1, 
         2200 Noordwijk, Netherlands \\
        E-mail: \email{gdemarchi@esa.int}}

\author{Nino Panagia\\
        Space Telescope Science Institute, 3700 San Martin Dr., Baltimore, MD 21218, USA\\
        INAF--NA, Osservatorio Astronomico di Capodimonte, Salita Moiariello
  16, 80131 Naples, Italy\\
Supernova Ltd, OYV \#131, Northsound Rd., Virgin Gorda VG1150,
  Virgin Islands, UK\\
        E-mail: \email{panagia@stsci.edu}}

\abstract{We have undertaken a systematic study of pre-main sequence (PMS) stars
spanning a wide range of masses ($0.5 - 4$\,M$_\odot$), metallicities
($0.1 - 1$\,Z$_\odot$) and ages ($0.5 - 30$\,Myr). We have used the {\em
Hubble Space Telescope} (HST)
to identify and characterise a large sample of PMS objects in
several star-forming regions in the Magellanic Clouds, namely 30\,Dor and
the SN\,1987A field in the LMC, and NGC\,346 and NGC\,602
in the SMC, and have compared them to PMS stars in similar
regions in the Milky Way, such as NGC\,3603 and Trumpler\,14, which we
studied with the HST and {\em Very Large Telescope} (VLT).

We have developed a novel method that combines broad-band
($V$, $I$) photometry with narrow-band $H\alpha$ imaging to determine the
physical parameters (temperature, luminosity, age, mass
and mass accretion rate) of more than 3000 bona-fide PMS stars still
undergoing active mass accretion. This is presently the largest and most
homogeneous sample of PMS objects with known physical properties and
includes not only very young objects, but also PMS stars older than
$10 - 20$\,Myr that are approaching the main sequence (MS).

We find that the mass accretion rate scales roughly with the square root
of the age, with the mass of the star to the power of 1.5, and with
the inverse of the cube root of the metallicity. The mass accretion rates
for stars of the same mass and age are thus systematically higher in the
Magellanic Clouds than in the Milky Way. These results are bound to have
important implications for, and constraints on our understanding of the
star formation process.

}

\FullConference{Frontier Research in Astrophysics\\
		 26 -- 31 May, 2014\\
		 Mondello (Palermo), Italy}

\begin{document}

\section{Introduction}

In the currently accepted star formation scenario (e.g. Lynden-Bell \& 
Pringle 1974; Bertout 1989), low-mass stars grow over time through 
accretion of matter from a circumstellar disc. A reliable measurement of the 
rate at which mass is accreted onto PMS stars is of paramount importance for 
understanding the evolution of both the stars and their discs (Calvet et al. 
2000). The formation of planetary systems is intimately connected with the 
properties of the circumstellar discs in which they are born (Wolf et al. 
2012). In particular, the timescale of disc survival sets 
an upper limit on the timescale of planet formation, becoming a stringent 
constraint for planet formation theories (e.g. Haisch, Lada \& Lada 2001). 
Therefore, it is particularly important to determine how 
the mass accretion rate varies with time as a star approaches the MS, how it 
depends on the mass of the forming star and how it is affected by the 
chemical composition and density of the parent molecular clouds or by the 
proximity of massive stars.

Observations of nearby star forming regions have reported that direct and 
indirect indicators of inner discs seem to disappear rapidly in the first 
few Myr of PMS evolution: both the fraction of young stars with infrared 
excess and those with evidence of disc accretion have been found to drop 
exponentially with the age of the stellar populations, with an e-folding 
time of $2-3$\,Myr (Hernandez et al. 2007; Williams \& Cieza 2011; Fedele 
et al. 2010). Concerning mass accretion, ground-based spectroscopic 
studies of nearby young star-forming regions (e.g. Taurus, Auriga, 
Ophiuchus; e.g. Sicilia--Aguilar et al. 2006) show that the mass accretion 
rate appears to decrease steadily with time, from about 
$10^{-8}$\,M$_\odot$\,yr$^{-1}$ at ages of $\sim 1$\,Myr for stars of about 
$0.5$\,M$_\odot$ to less than $10^{-9}$\,M$_\odot$\,yr$^{-1}$  for the same 
stars at ages of $\sim 10$\,Myr (e.g. Muzerolle et al. 2000; 
Sicilia--Aguilar et al. 2005; 2006; 2010). At face value this is in line 
with the expected evolution of viscous discs (Hartmann et al. 1998), 
even though the scatter of the data exceeds 2 dex at any given age (see also 
Figure\,\ref{fig3}), also because the observations cover a wide and sparsely 
populated mass range.

Extrapolating from these nearby results, one would be tempted to conclude
that circumstellar discs are rather short lived and star formation is a 
rather quick process. Unfortunately, all the studies mentioned above address 
a very limited sample of star formation properties. All regions studied so 
far are nearby; they contain few stars, typically with masses of a few 
tenths of M$_\odot$; they are not in dense environments; they do not 
contain massive stars, except for Orion; and they all have solar 
metallicity. Thus, these environments are not representative of massive 
starburst clusters, where possibly most stars form in galaxies. Nor are 
they characteristic of the conditions in place when most stars formed in the 
universe, at redshift $z \gtrsim 2$ (e.g. Madau et al. 1996; Lilly et 
al. 1996), when metallicity was about $1/3$\,Z$_\odot$. For this reason, it 
is important to study star formation in other places in our Galaxy, not only 
around the Sun but also in massive clusters, and also in nearby Galaxies, 
for instance the Magellanic Clouds (MCs). With a metallicity between 
$1/3$\,Z$_\odot$ and $1/10$\,Z$_\odot$, the MCs offer us an environment 
with the prevailing conditions at $z \simeq 2$.

Of course, observing star-forming regions farther away than the solar 
neighbourhood introduces some practical problems, mostly due to angular 
resolution. For instance, while multi-object spectrographs at large ground 
based observatories are sensitive enough to give us good spectra of PMS 
stars in the MCs, atmospheric seeing makes these studies extremely difficult 
for all but the most massive PMS objects (e.g. Kalari et al. 2014). 
Alternatively, the the properties of circumstellar discs around
PMS stars in the MCs can be studied with infrared space observatories such 
as {\em Spitzer} and {\em Herschel} (e.g. Seale et al. 2009; Carlson et al. 
2011; Carlson et al. 2012; Meixner et al. 2013). However, also in this
case angular resolution remains a problem and all sources detected in
this way and classified as young stellar objects (YSOs) are in fact  
groups of YSOs or even small clusters and the properties of individual 
stars cannot be derived. 

This investigation will benefit tremendously from the {\em James Webb Space 
Telescope} (JWST; Gardner et al. 2006) and in particular 
from NIRSpec, the multi-object spectrograph provided by the European Space 
Agency (Birkmann et al. 2014). NIRSpec will deliver
simultaneous near-infrared spectra ($0.6 - 5.0$\,$\mu$m) of up to $\sim 100$ 
PMS stars wherever located inside its $3^\prime \times 3^\prime$ field of 
view, with a spatial resolution equivalent to that of the HST at optical 
wavelengths ($< 0.1"$). The strength of the accretion process and 
mass accretion rates can be readily derived from the analysis of prominent 
recombination lines in the Paschen and Brackett series (e.g. Natta, Testi
\& Randich 2006). 

Already now, however, there is a very efficient and powerful way to obtain
this information, for hundreds of stars simultaneously, from photometry 
alone. Using the HST and VLT, in collaboration with a group of European 
colleagues (see Acknowledgments) we have started a study of the PMS phase in 
a number of star forming regions in the local group. These include NGC\,3603
and Trumpler\,14 in the Milky Way, 30\,Doradus and surrounding regions in 
the Large Magellanic Cloud (LMC), and NGC\,346 and NGC\,602 in the Small 
Magellanic Cloud (SMC). In the following sections we describe the 
method and its application and present the most important results of this
investigation so far.

\section{The method}

The method hinges on the fact that the spectra of PMS stars undergoing 
mass accretion have distinctive emission features, in particular strong 
emission in the $H\alpha$ line (equivalent width $W_{\rm eq} > 10$\,\AA), 
due to the recombination of the gas that
is shocked and ionised in the accretion process. Thanks to these features,
we have shown that it is possible to efficiently and reliably
identify all objects of this type in a stellar field, regardless of
their age and of their position in the colour--magnitude diagram (CMD). 
Building on the work of Romaniello (1998), of Panagia et al. (2000), and 
of Romaniello et al. (2004), in De Marchi, Panagia \& Romaniello (2010) 
and De Marchi et al. (2011a), we 
showed that a suitable combination of broad- and narrow-band photometry 
provides an accurate determination of the $H\alpha$ luminosity of these 
objects. From it, the accretion luminosity and mass accretion rate can
easily be derived. 

\begin{figure*}
\centering
\includegraphics[width=0.32\textwidth,height=5cm,trim=30 10 30 30]
{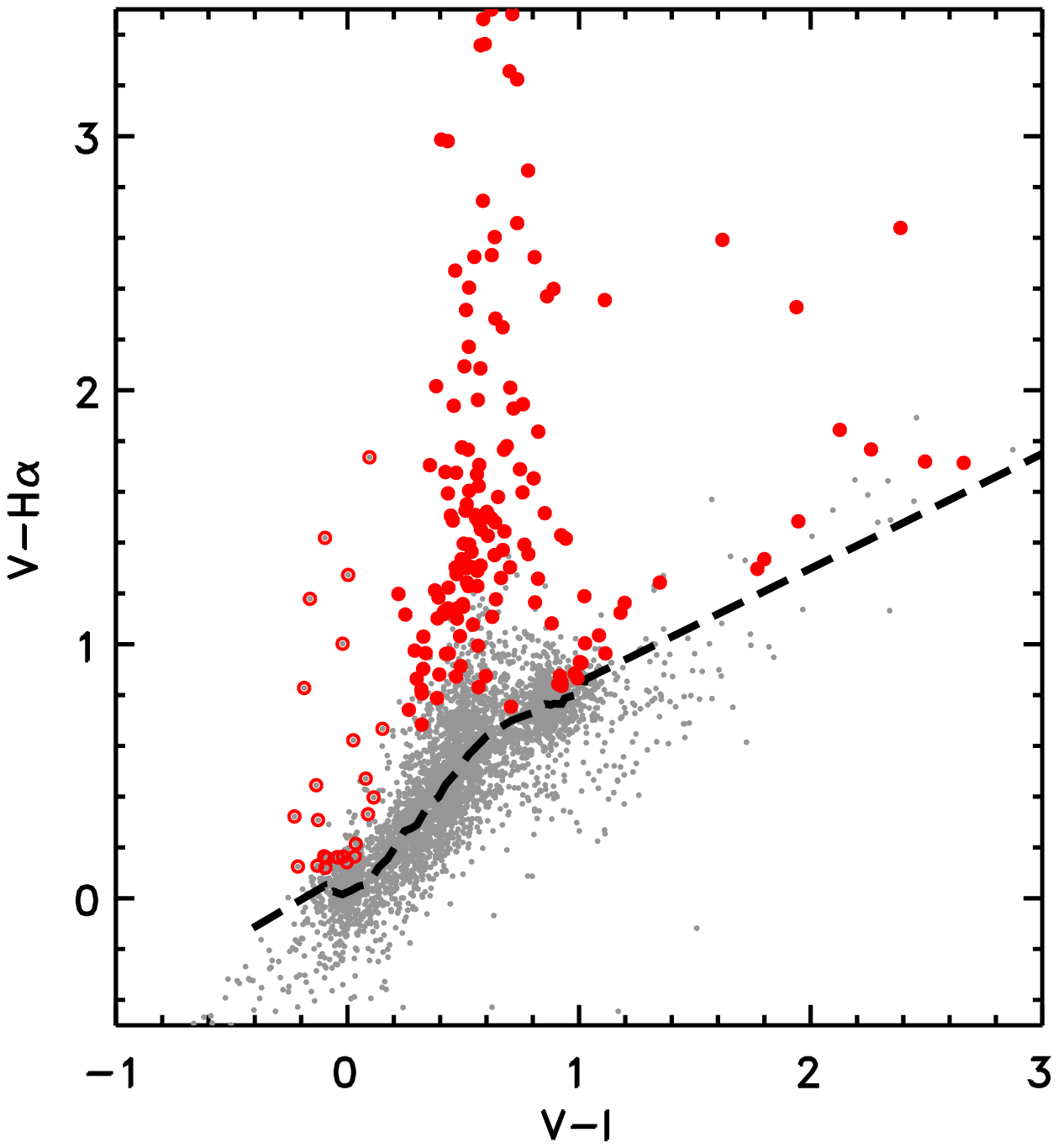}
\includegraphics[width=0.32\textwidth,height=5cm,trim=30 10 30 30]
{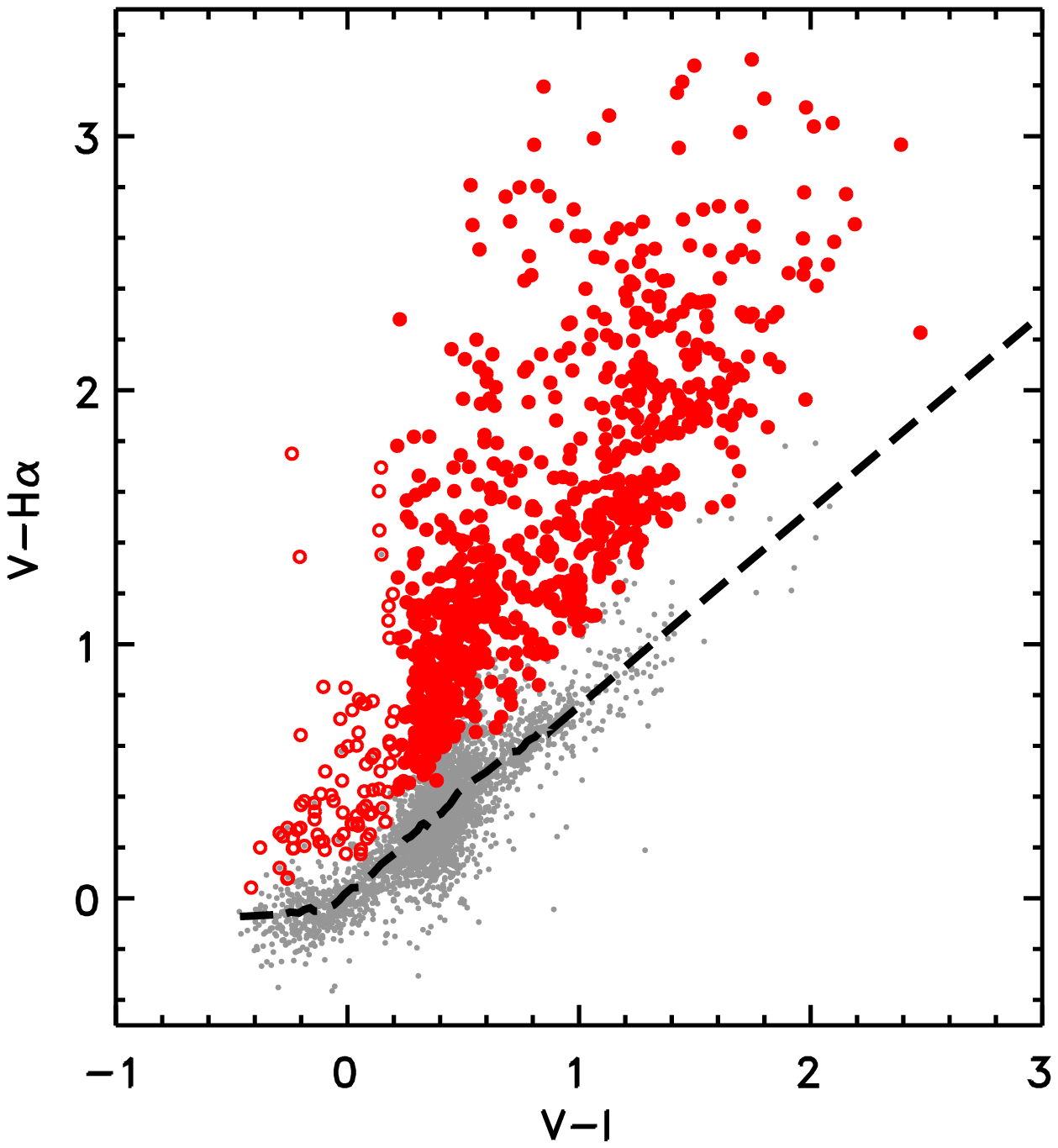}
\includegraphics[width=0.32\textwidth,height=5cm,trim=30 10 30 30]
{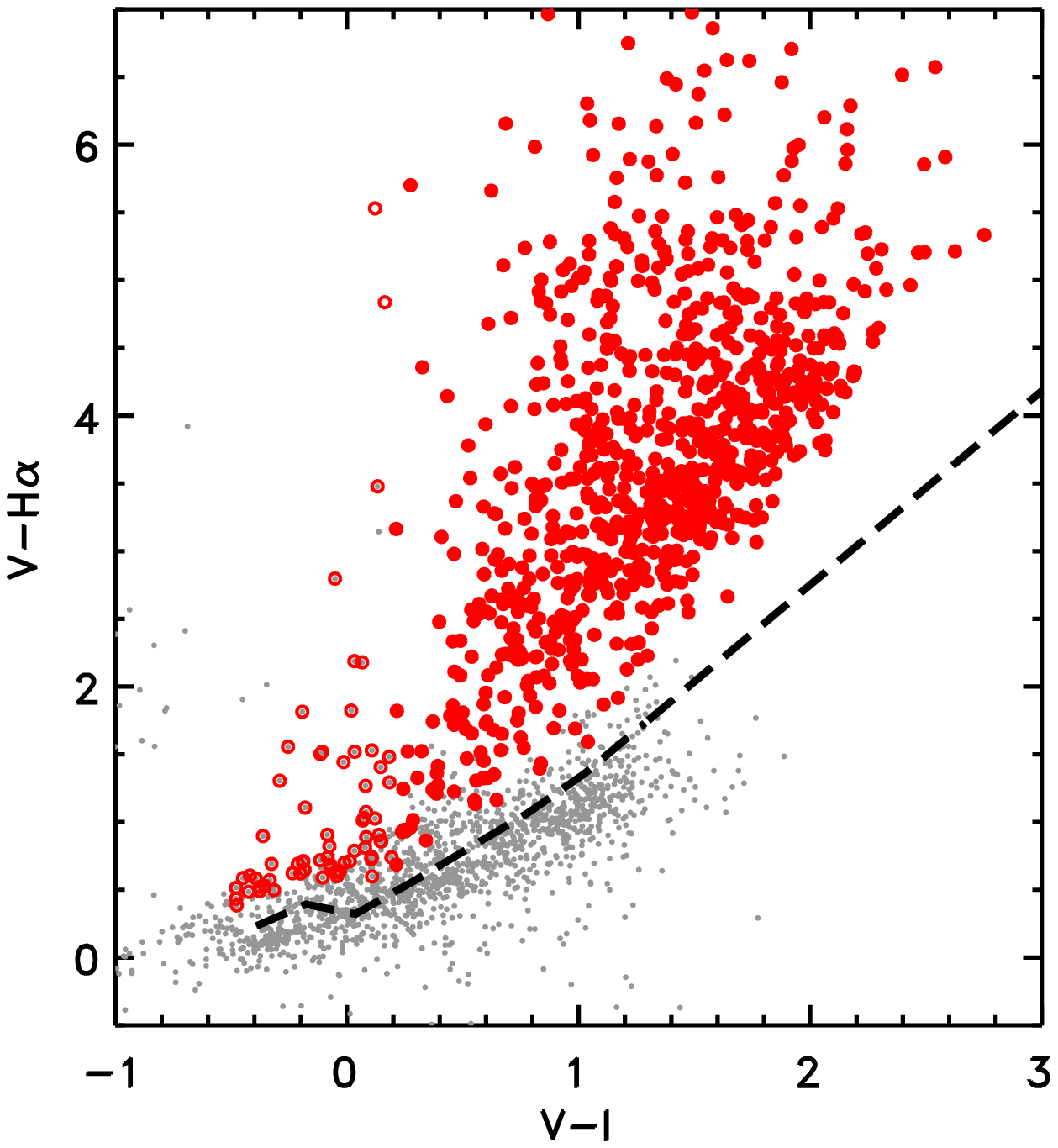}
\caption{Run of the $V-H\alpha$ colour as a function of $V-I$ in a field
around SN\,1987A in the LMC (left; De Marchi et al. 2010) in the
cluster  NGC\,346 in the SMC (middle; De Marchi et al. 2011a) and in
NGC\,3603 in the Galaxy (right; Beccari et al. 2010). All colours are
corrected for extinction.}
\label{fig1}
\end{figure*}

We look for the presence of an excess in the $H\alpha$ emission line by 
using a combination of broad-band ($V$, $I$) and narrow-band ($H\alpha$) 
photometry. This way of identifying PMS stars is more reliable than the 
simple classification based on the position of the objects in the CMD or 
Hertzsprung--Russell (H--R) diagram and provides a secure detection of 
relatively old PMS stars, already close to the MS.

The virtue of this new method is that it derives the luminosity of the 
photospheric continuum of a star inside the specific $H\alpha$ band simply 
by interpolation from the average $V-H\alpha$ colour of stars with the 
same $V-I$ index (see Figure\,\ref{fig1}). As shown in De Marchi et al. 
(2010), the majority of stars 
in a typical stellar field have no excess $H\alpha$ emission. Therefore, the 
{\em median} value of the $V - H\alpha$ colour index at a given effective
temperature $T_{\rm eff}$ defines a spectral reference template for all
stars with that $T_{\rm eff}$ and can be used to identify objects with
$H\alpha$ excess. Equipped with the knowledge of the continuum level in the 
$H\alpha$ band, we can easily determine the $H\alpha$ luminosity, 
$L(H\alpha)$, of each star. 

The method and its applications are fully explained in a series of papers 
(De Marchi et al. 2010, 2011a, 2011b, 2011c, 2013a; 
Beccari et al. 2010; Spezzi et al. 2012; Beccari et al. 2015) and the 
accuracy of the $H\alpha$ continuum and $L(H\alpha)$ derived in this way 
has been independently confirmed with spectroscopic measurements by 
Barentsen et al. (2011).

Some examples of how the method works are presented in Figure\,\ref{fig1}, 
where we show the run of the $V-H\alpha$ colour as a function of $V-I$ in 
three fields. From left to right, CMDs correspond to stars in the field 
around SN\,1987A in the LMC, in the cluster NGC\,346 
in the SMC, and in NGC\,3603 in the Galaxy. Stars with small photometric 
uncertainties (grey dots) define the reference template for 
normal stars (i.e. with no $H\alpha$ emission), shown as a dashed line (the 
lines are not the same in the three panels due to differences between the 
different HST cameras used for the observations). When $H\alpha$ emission is 
present in PMS stars, this results in a greater than average value of the 
$V-H\alpha$ colour. We conservatively take as PMS objects all stars with 
$V-H\alpha$ colour departing from the template line more than four times 
their photometric uncertainty (thick red dots). We have identified in this 
way more than 3\,400 bona-fide PMS stars, in the three galaxies, still 
undergoing mass accretion. We are currently extending the sample with the
addition of about 10\,000 PMS objects (De Marchi, Panagia, Sabbi 2015, in
preparation) revealed with our method in a region of $\sim 14^\prime \times
12^\prime$ centered on 30\,Dor, observed as part
of the Hubble Treasury Tarantula Survey (Sabbi et al. 2013).

\begin{figure*}[ht]
\includegraphics[width=0.49\textwidth,height=7cm,trim=30 10 30 30]
{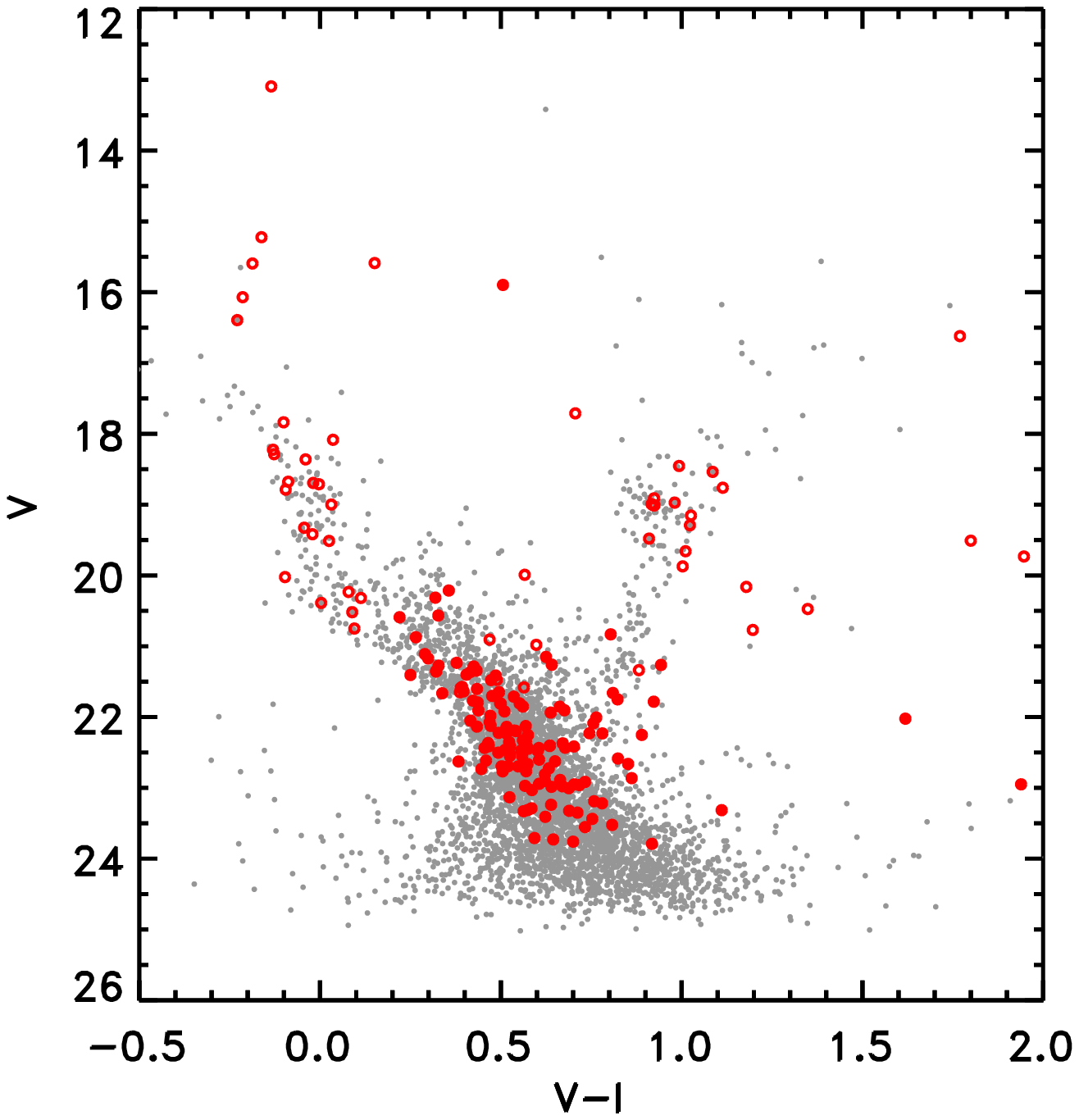}
\hfill
\includegraphics[width=0.49\textwidth,height=7cm,trim=30 10 30 30]
{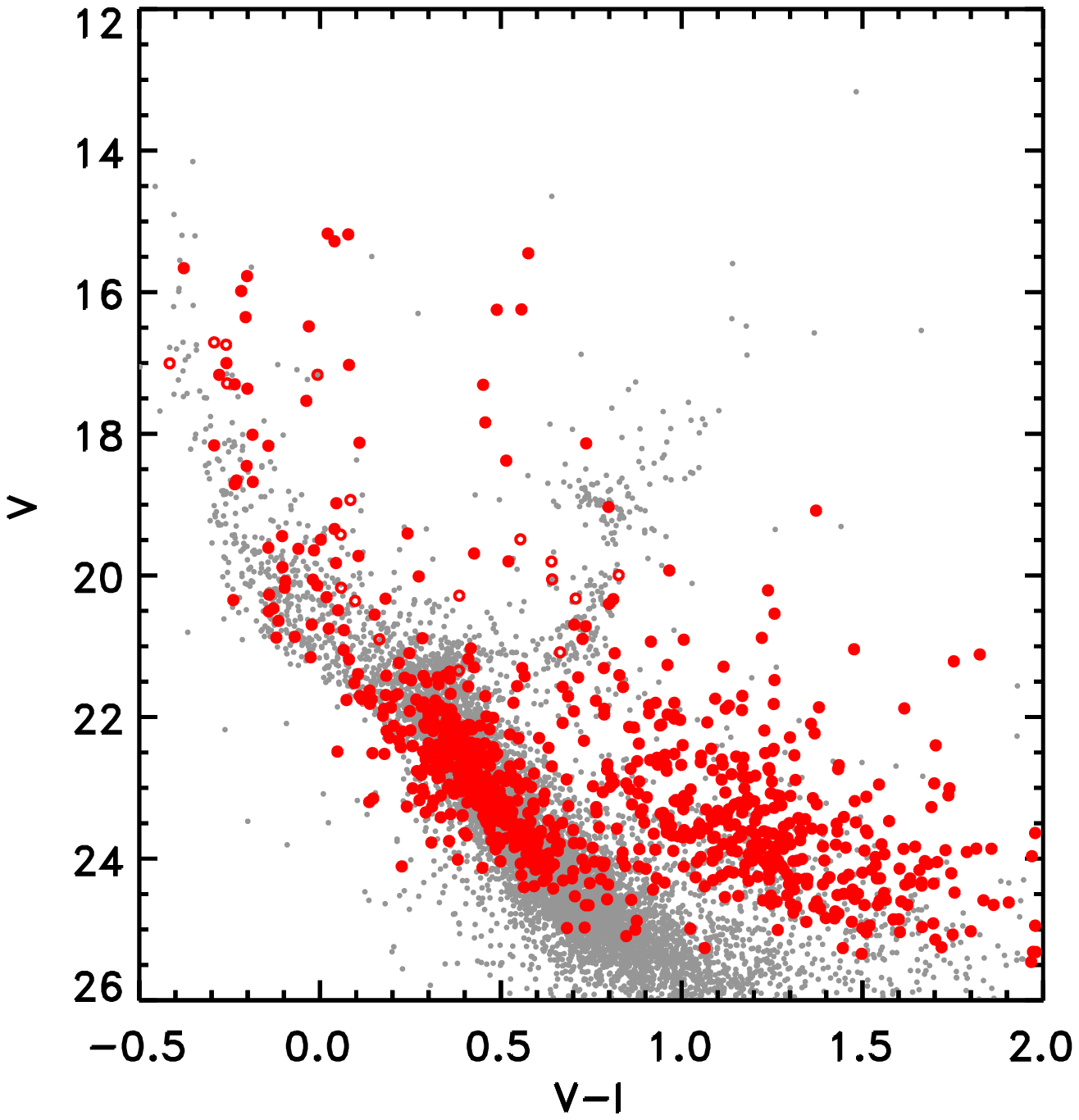}
\caption{Colour--magnitude diagrams of all stars in the field around
SN\,1987A in the LMC (left), and in the one containing the cluster NGC\,346 
in the SMC (right). Stars with
an $H\alpha$ excess identified as in Figure\,1 are highlighted. Whilst
some are still very young, and as such still very distant from the MS,
many are older objects already approaching the MS, and as such would not
be distinguishable from normal MS stars using standard broad-band
photometry alone. By detecting all stars with an $H\alpha$ excess, we
can easily identify multiple stellar populations in the same region and
study separately their physical properties (spatial location, age, mass,
mass accretion rate).}
\label{fig2}
\end{figure*}

\section{Multiple generations}

The method that we have developed to detect PMS stars in a stellar field 
allows us to identify {\em all} objects with an excess emission. This means
that we are able to detect not only the most recent generation of T Tauri 
stars, whose colours are still dominated by the circumstellar discs, but also  
relatively ``mature'' PMS stars, already close to the MS. An example is 
shown in Figure\,\ref{fig2}, where all objects indicated by red dots are stars 
currently displaying $H\alpha$ excess emission due to the ongoing active 
mass accretion. Broad-band photometry alone could not distinguish the bluer
objects (i.e. those near the MS) from normal MS stars, but the $H\alpha$
photometry allows us to securely identify them as still intrinsically 
young. 

Comparison of the CMDs of Figure\,\ref{fig2} with theoretical isochrones 
immediately
suggests an older age for the bluer objects, typically older than $\sim 
10$\,Myr. There are, of course, some uncertainties on the ages derived 
through isochrone comparison: besides photometric errors and uncertainties 
in the input physics affecting the models, there are other physical
effects that could cause an incorrect determination of the age (or mass) of 
individual objects. These include for instance unresolved binaries, 
differential reddening, stellar variability, veiling resulting from 
accretion, and scattering due to a disc seen at high inclination. All 
these effects can combine to produce a broadening in the CMD, which could 
be misinterpreted as an age spread (see e.g. Hennekemper et al. 2007
and Da Rio et al. 2010).

\begin{figure*}[t]
\includegraphics[width=0.49\textwidth,height=6.5cm,trim=30 10 30 30]
{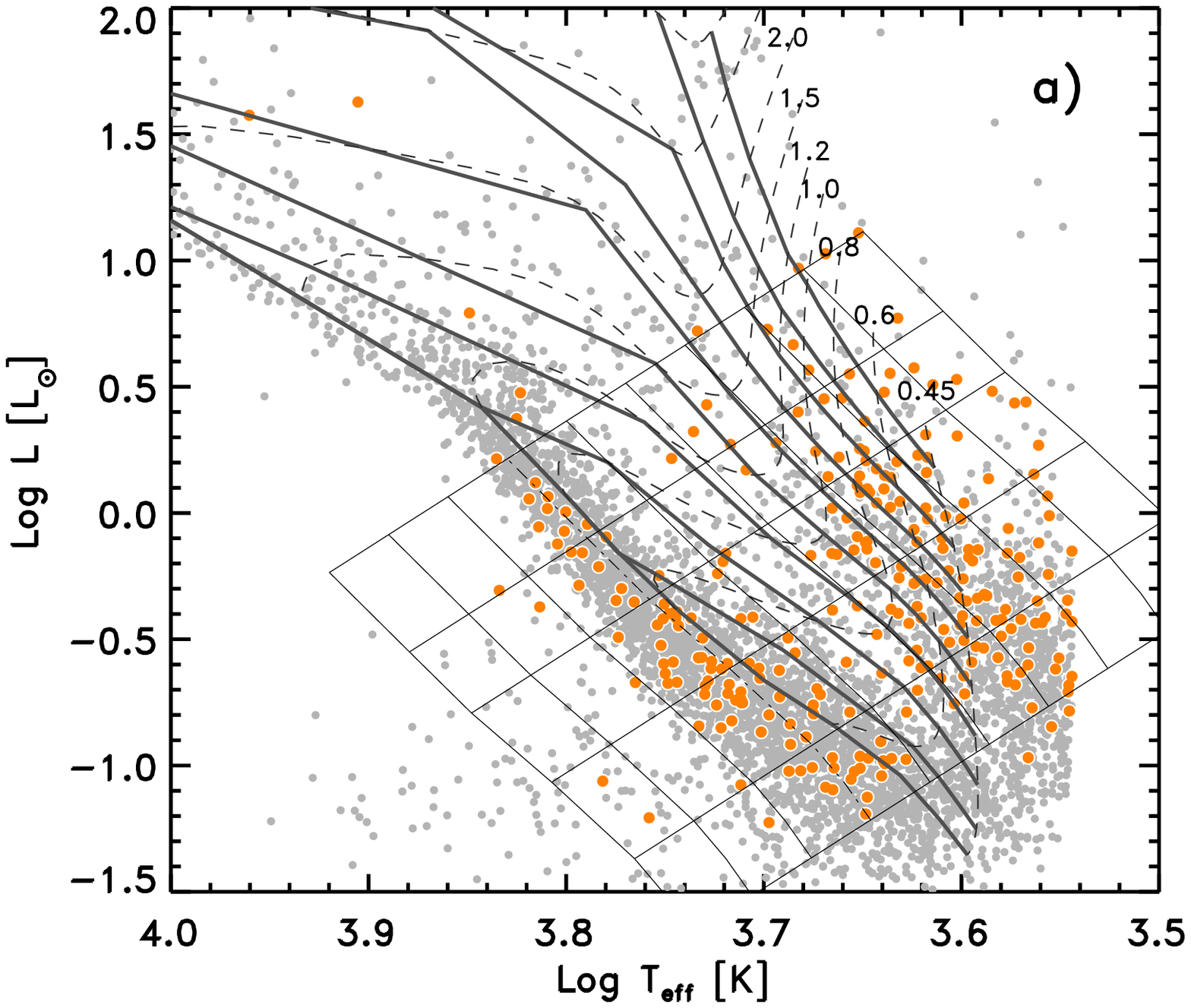}
\hfill
\includegraphics[width=0.49\textwidth,height=6.5cm,trim=30 10 30 30]
{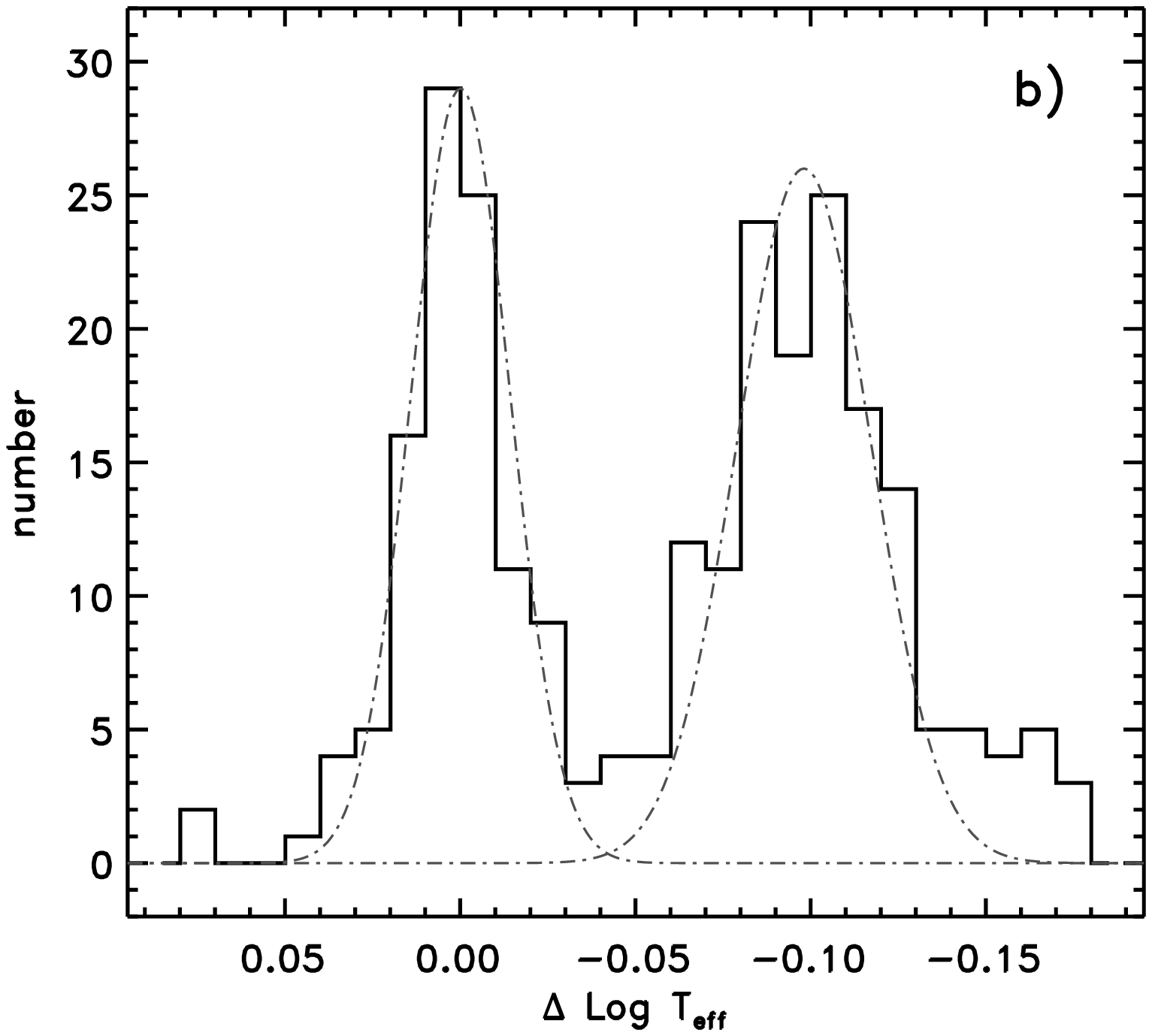}
\caption{Distribution of PMS stars in the H--R diagram of NGC\,602, 
obtained by counting the number of objects falling in strips parallel to the 
ZAMS. Thick solid lines show the evolutionary tracks for metallicity $Z = 
0.004$ and masses as indicated. The corresponding isochrones are shown as 
thin dashed lines, for ages of $0.125, 0.25, 0.5, 1, 2, 4, 8, 16, 32,$ and 
64\,Myr from right to left.
The grid shown in panel a) corresponds to $\log T_{\rm eff} = 0.03$, 
for display purposes, but the histograms in panel b) are obtained using a 
finer grid ($\log T_{\rm eff} = 0.01$). The value of $\log T_{\rm eff} = 0.$
corresponds to the dot-dashed line in panel a). The thin dot-dashed lines in 
panel b) show a Gaussian fit to the two peaks.}
\label{fig3}
\end{figure*}

In fact, even though all the effects mentioned above could mimic an age 
spread, none of them can produce the clearly bimodal distribution that we 
observe in the CMDs. We show this in a quantitative way in Figure\,\ref{fig3}, 
containing in the left panel the H--R diagram of NGC\,602 (De Marchi, 
Beccari \& Panagia 2013). The remarkable paucity of PMS stars with age 
comprised between the 4\,Myr and 16\,Myr isochrones is quantified in the 
right panel, where we show the distribution of PMS stars as a function of 
the difference in their $\log T_{\rm eff}$ values. The distribution is 
obtained by counting the number of PMS stars in strips parallel to the 
zero-age MS, starting from the dot-dashed line, and it is clearly 
bimodal, with two peaks separated by several times their width. A Gaussian
fit to the two peaks (dot-dashed lines in Figure\,\ref{fig3}b) gives 
$\sigma_1 = 
0.020$\,dex for older PMS stars and $\sigma_2 = 0.025$\,dex for younger PMS 
stars. The separation between the two peaks ($0.1$\,dex) corresponds to
respectively 5 and 4 times these widths and confirms that the two 
distributions are clearly distinct.

Physically, it is hard to imagine a mechanism that would affect the 
temperatures and luminosities of stars in such a way that they are 
selectively displaced from the region occupied by young PMS objects in the 
H--R diagram and moved towards the MS, while leaving only a handful of them 
in the region in between. Therefore, we can safely conclude that the
two groups of stars with $H\alpha$ excess seen in the CMDs must belong to 
different generations, with ages that differ by much more than a factor of 
two and likely up to an order of magnitude.

Besides very different ages, the two populations of younger and older PMS 
stars shown in Figures\,\ref{fig2} and \ref{fig3} also have considerably 
different spatial
density distributions. As an example, for NGC\,602 we compare these 
distributions to one another in Figure\,4 by means of contour lines of 
stellar density with logarithmic steps, overlaid on a true-colour image
of the region. We have selected all stars younger than 5 Myr (cyan) and 
those older than 20 Myr (orange). The lowest contour level corresponds to a 
local 
density of PMS stars three times as high as the average PMS stars density 
over the entire field. The steps between contour levels are constant and 
corresponds to 0.3\,dex. We also show with yellow dots the 
positions of the few stars (34 in total) with ages between 5 and 20 Myr. 

\begin{figure*}[ht]
\centering
\includegraphics[width=0.8\textwidth,height=11.5cm,trim=30 10 30 30]
{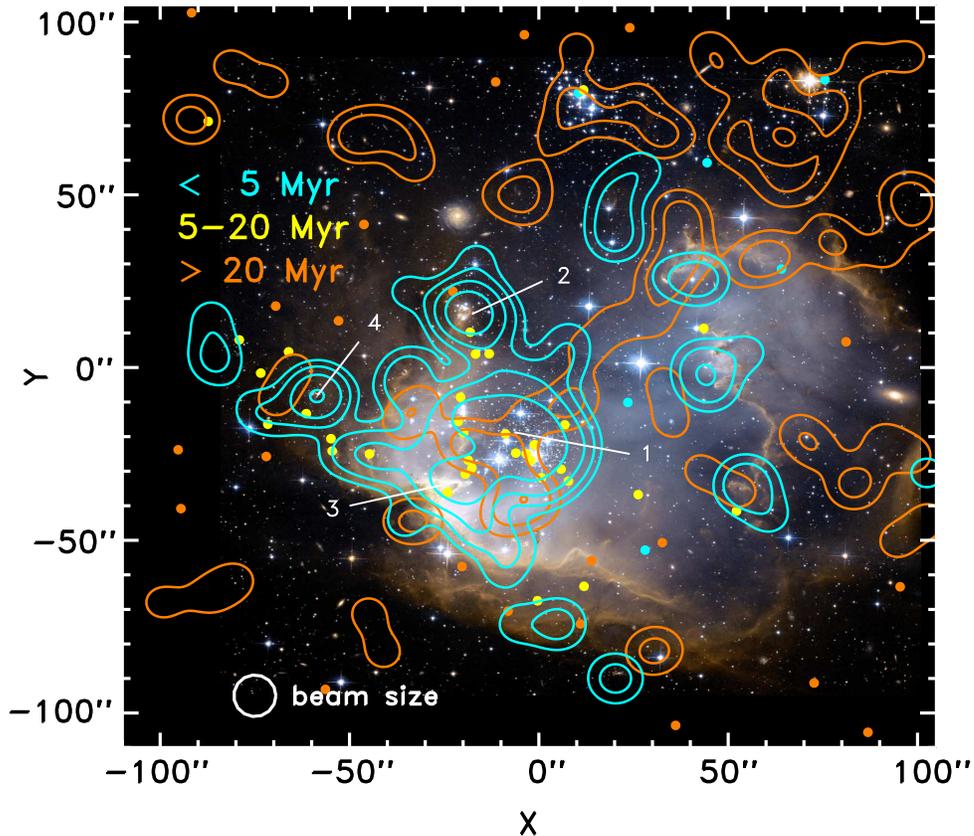}
\caption{Contour plots showing the position and density distribution of 
PMS stars of different ages, as per the legend, overlaid on a true-colour 
image of NGC\,602. The lowest level corresponds to a density three times 
as high as the average density of PMS stars in the field. The
actual stellar distributions are convoluted with the beam of $\sigma
= 4^{\prime\prime}$, or $\sim 1.2$\,pc, shown in the figure. The step 
between contour levels is constant and corresponds to a factor of 2.}
\label{fig4}
\end{figure*}

The spatial distribution of younger and older PMS stars are remarkably 
different: older objects are much more widely distributed and, except for 
the centre, they do not always overlap with the younger generation. 
This confirms that episodic accretion (e.g. Baraffe et al. 2009, 2010) is
not at the origin of the bimodal distribution of PMS stars as seen in 
Figure\,\ref{fig3}, because the effects of episodic accretion could move 
stars from one
side to the other of the CMD, but they cannot possibly move the stars across
the sky! Hence we must conclude that there is a fundamental difference
between the properties of bluer and redder PMS stars in the CMD, and their
age is the difference. This finding is not limited to NGC\,602, but is 
a common feature of all the star-forming regions that we have investigated 
so far in the three galaxies.

In summary, all regions that we studied in the Milky Way, LMC and SMC 
exhibit multiple recent episodes of star formation, indicating that star 
formation has proceeded over a long
time, even though our age resolution cannot discriminate between an
extended episode or short and frequent bursts. We also find that there
is no correlation between the projected spatial distribution of young
and old PMS stars and that the younger population is systematically more
concentrated.

\begin{figure}[ht]
\centering
\includegraphics[width=0.49\textwidth,height=6.5cm]
  {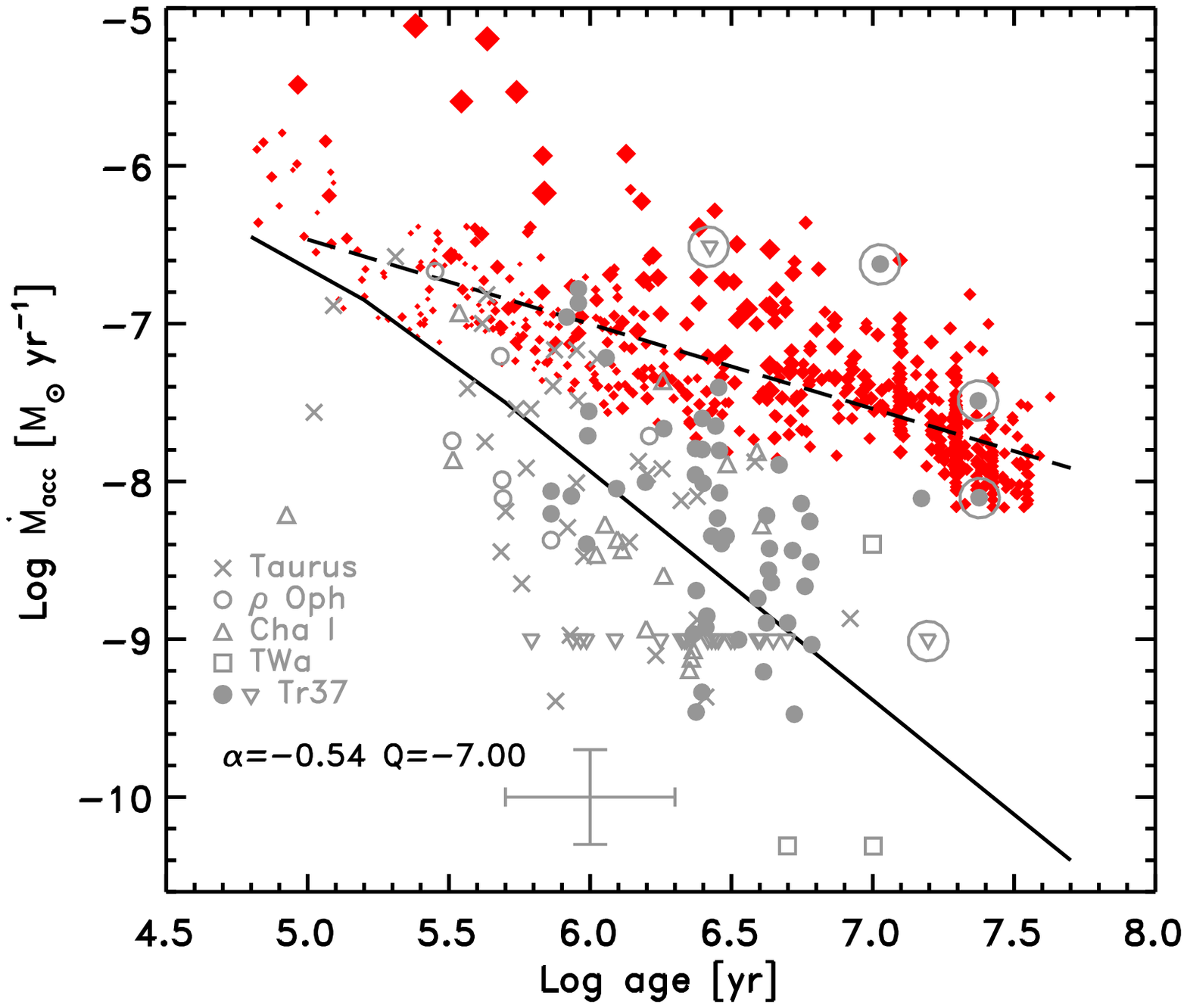}
\hfill
\includegraphics[width=0.49\textwidth,height=6.5cm,bb=7 360 560 850]
  {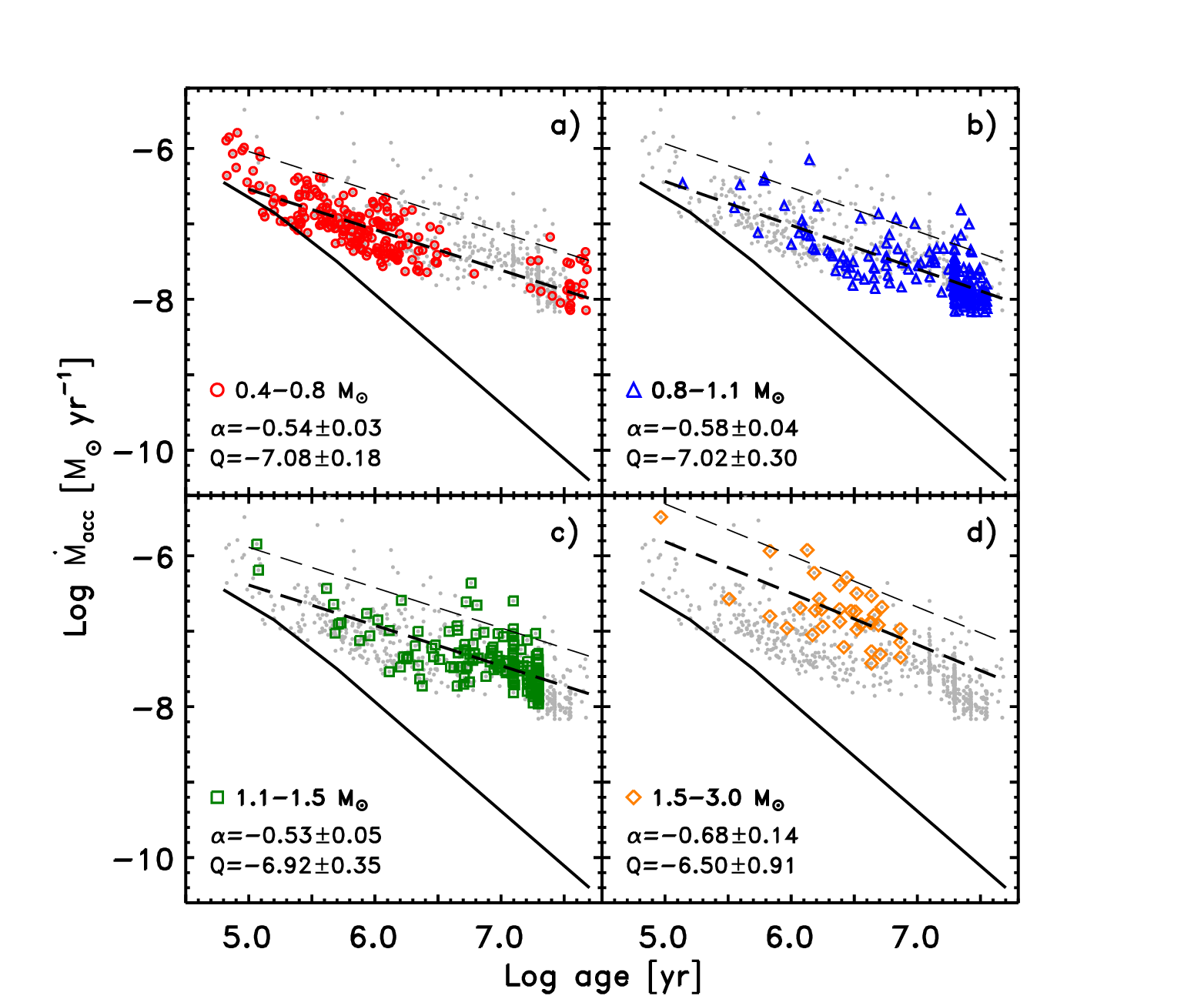}
\vspace{-0.3cm}
\caption{The left panel shows the mass accretion rate as a function of
stellar age for PMS stars in NGC\,346 (diamonds) compared with that of
Galactic T Tauri stars (see legend) from the work of Sicilia--Aguilar et
al. (2006; the large cross indicates the uncertainties as quoted in that
paper). The solid line shows the current models of viscous
disc evolution from Hartmann et al. (1998). Our measurements are
systematically higher than the models, and the effect remains when we
consider separately stars of different masses (right panel, see legends
for mass values). The right panel shows that all four mass groups have 
the same decline of $\dot M_{\rm acc}$ with age ($\alpha \simeq -0.5$, 
thick dashed lines), but $\dot M_{\rm acc}$ is higher for more massive 
stars (see value of intercept $Q$ at 1\,Myr).}
\label{fig5}
\end{figure}

\section{Evolution of the mass accretion rate}

Using the complete set of multi-band HST photometry available for these 
objects, and through comparison with evolutionary tracks for the appropriate
metallicities, we have determined their physical parameters, including 
temperature, luminosity, age, mass and mass accretion rate. This is 
presently the largest and most homogeneous sample of PMS objects with 
known physical properties. 

A fundamental parameter that we can derive with this method is the mass
accretion rate, $\dot M_{acc}$. Since the energy released by the
accretion process goes towards ionising and heating the circumstellar
gas, the accretion luminosity $L_{acc}$ can be derived from 
$L_{H\alpha}$. With the mass and radius of each PMS star determined from
the evolutionary tracks, the value of $\dot M_{acc}$ can be obtained
from the free-fall equation, linking the luminosity released by
the impact of the accretion flow with the rate of mass accretion,
according to the relationship:

\begin{equation}
L_{\rm acc} \simeq \frac{G\,M_*\,\dot M_{\rm acc}}{R_*} \left(1 -
\frac{R_*}{R_{\rm in}}\right)
\label{eq1}
\end{equation}

\noindent
where $G$ is the gravitational constant, $M_*$  the mass of the star
determined above, $R_*$ its photospheric radius coming from its
luminosity and effective temperature, and $R_{\rm in}$ the inner radius
of the accretion disc. The value of $R_{\rm in}$ is rather uncertain and
depends on how exactly the accretion disc is coupled with the magnetic
field of the star. Following Gullbring et al. (1998), we adopt $R_{\rm
in} = 5\,R_*$ for all PMS objects and with this assumption we have all
the parameters needed to determine $\dot M_{\rm acc}$.

We find that older PMS stars have typically lower mass accretion rates, as
shown for instance in the left panel of Figure\,\ref{fig5} relative to the
region of NGC\,346 in the SMC. The long dashed line represents the best fit 
to the observed distribution of mass accretion rates and its slope
($\alpha=-0.55$) is rather similar to the one measured for stars of similar 
masses and ages in NGC\,602 ($\alpha=-0.7$; De Marchi et al. 2013). 

On the other hand, both slopes are considerably shallower than the 
$\sim t^{-1.5}$ decline predicted by the models of Hartmann et al. 
(1998; see also Calvet et al. 2000; Muzerolle et al. 2000) for 
viscous disc evolution, represented here by the solid line. The latter 
appears to reproduce rather well the trend of decreasing $\dot 
M_{\rm acc}$ with stellar age for low-mass Galactic T-Tauri stars as 
compiled by Sicilia--Aguilar et al. (2006). At first glance, this discrepancy 
could seem to imply a different evolution of the mass accretion rate for 
PMS stars in the Galaxy and in the SMC. However, before any meaningful 
comparison can be made
the masses of the individual objects must be taken explicitly into account.
In fact, there is a marked difference in the mass ranges covered by our
observations and those in the compilation of Sicilia--Aguilar et al. (2006).
The latter includes very-low mass nearby objects, typically a few tenths of 
M$_\odot$, while our much more distant objects are 
typically more massive than $0.5$\,M$_\odot$. 


The large size of our sample of PMS stars allows us to study the mass 
dependence of $\dot M_{\rm acc}$ in a robust way. We show this graphically 
for NGC\,346 in the right-hand panel of Figure\,\ref{fig5}, where we have 
split our sample in four roughly equally populated mass groups, namely 
$0.4 - 0.8$\,M$_\odot$, $0.8 - 1.1$\,M$_\odot$, $1.1 - 1.5$\,M$_\odot$, and 
$1.5 - 3.0$\,M$_\odot$, and show for each one separately the run of $\dot 
M_{\rm acc}$ as a function of age. Each panel gives the slope $\alpha$ and 
intercept $Q$ (at 1\,Myr) of the best linear fit to the data (thick 
long-dashed lines), according to the relationship $\log \dot M_{\rm acc} = 
\alpha \times \log({\rm age}) + Q$ with the age in Myr. These values are
in excellent agreement with those found in NGC\,602 (De Marchi et
al. 2013) in the same galaxy, as mentioned above.

The figure shows that all four mass groups have
the same decline of $\dot M_{\rm acc}$ with age ($\alpha \simeq -0.5$,
thick dashed lines), but $\dot M_{\rm acc}$ is higher for more massive
stars (see value of intercept $Q$ at 1\,Myr). Although our method will
inevitably miss some stars with weak $H\alpha$ excess emission, the
upper envelopes of the distribution (thin dashed lines) appear to be
fully consistent with the slope of the best fit.

Since our sample is quite rich, we can go a step further and perform a 
multivariate least-square fit to the observations to derive the simultaneous
dependence of $\dot M_{\rm acc}$ on both the stellar mass and age. We 
assume a relationship of the type: 

\begin{equation}
\log \dot M_{\rm acc} = a \times \log t + b \times \log m + c,
\label{eq2}
\end{equation}

\noindent
where $t$ is the age in Myr, $m$ the mass in solar units and $c$ a constant, 
corresponding to the intercept at 1\,Myr and 1\,M$_\odot$, (note that $c$ is 
similar to the parameter $Q$ defined above, but it characterises the 
simultaneous fit on mass and age). The resulting best fit gives $a=-0.59 
\pm 0.02$ and $b=0.82 \pm 0.09$, while for NGC\,602 the same parameters are 
$a=-0.72 \pm 0.02$ and $b=0.94 \pm 0.14$, confirming a rather similar
dependence on mass and age for SMC stars.

Our observations have already shown that, at a given mass or age, the mass 
accretion rate appears to be higher in the SMC than in the LMC, and in 
turn higher in the LMC than in the Galaxy, suggesting that the mass accretion 
rate could also depend on the metallicity of the star-forming regions. To 
explore the nature of this trend, it is necessary to understand whether a 
relationship exists between the parameter $c$ and metallicity. Indeed, $c$ 
is in practice an ''effective'' mass accretion rate: inverting 
Equation\,\ref{eq2} one sees that, as mentioned above, $c$ is the mass 
accretion rate of a star with mass 1\,M$_\odot$ and age 1\,Myr.
 
A preliminary comparison of NGC\,346 and NGC\,602 with the results that
we obtain in the SN\,1987A field (De Marchi et al. 2010; Spezzi et al. 2012), 
in 30\,Dor (De Marchi et al. 2011c, and in preparation), of NGC\,3603 
(Beccari et al. 2010, and in preparation), and of Trumpler\,14 (Beccari et 
al. 2015) shows that using approximate values of $a=-0.5$ and $b=1.5$ still 
results in a satisfactory multivariate least-square fit. In other words, 
while the best-fitting relationship between mass accretion rate, stellar age 
and stellar mass for each individual region has different values of $a$ and 
$b$, we can constrain these parameters to take on the values of $a=-0.5$ and 
$b=1.5$ with still acceptably small residuals for all regions simultaneously. 
In this case, we can then look at the values of $c$ to explore the dependence 
of the mass accretion rate on the environment. 

\begin{figure} 
\centering
\includegraphics[width=0.5\textwidth,height=6cm,trim=30 10 30 30]{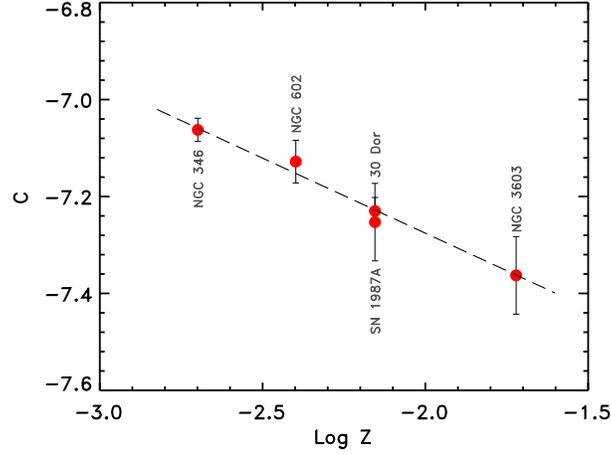}
\caption{Run of the ``effective'' mass accretion $c$ as a function of
the metallicity for the regions that we have studied so far (as indicated),
having assumed $a=-0.5$ and $b=1.5$ for all of them.
Although the uncertainties on the individual $c$ are not negligible, the
indication is rather clear that the mass accretion rate is higher in
environments of lower metallicity. The simple fit shown by the dashed line
corresponds to $c \propto Z^{-1/3}$.}
\label{fig6}
\end{figure}

The resulting values of $c$ are shown in Figure\,\ref{fig6} as a function 
of the metallicity $Z$ of each region, showing a rather remarkable 
correlation. A simple linear fit suggests that $c \propto Z^{-1/3}$. We can, 
therefore, rewrite Equation\,\ref{eq2} in an approximate form as:

\begin{equation}
\log \dot M_{\rm acc} \simeq -\frac{1}{2} \times \log t + 
 \frac{3}{2} \times \log m  - \frac{1}{3} \times \log Z - 7.9.
\label{eq3}
\end{equation}

These results are necessarily still preliminary, since they are based
primarily on observations obtained in dense regions of intense star 
formation. We have already secured HST observations of other more quiet 
and diffuse regions of star formation in all three these galaxies, 
in order to investigate how the properties of the environment affect 
the mass accretion process.

\begin{figure} 
\centering
\includegraphics[width=0.55\textwidth,height=8cm,trim=30 30 30 30]{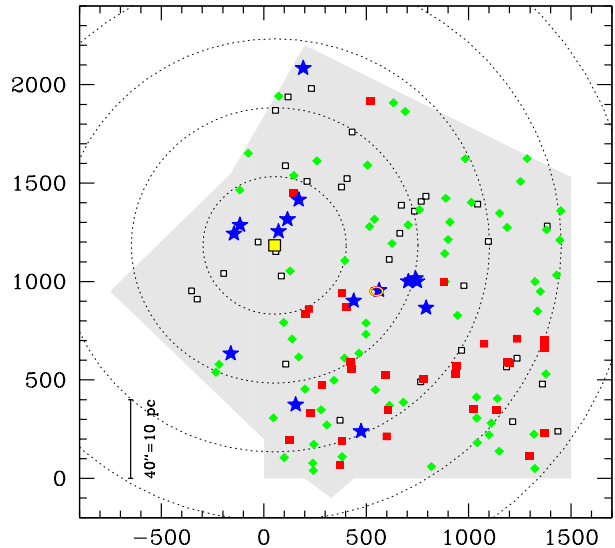}
\caption{Map of the distribution of all bona-fide PMS objects with respect
to the ionising stars (blue star symbols), whose barycentre is indicated
by the large yellow square.
For reference, a red ellipse marks the position of SN1987A.
Red squares correspond to stars with $L(H\alpha) >
2 \times 10^{-2}$\,L$_\odot$, white squares indicate objects with $L(H\alpha) <
8 \times 10^{-3}$\,L$_\odot$, and
green diamonds are used for intermediate values. Note the paucity of PMS
stars with high $L(H\alpha)$ near the ionising stars.}
\label{fig7}
\end{figure}

In fact, we have already traced the effects that nearby massive stars 
can have on the photoevaporation of circumstellar discs, even in the MCs.
For instance, in the field of SN\,1987A we find a clear anti-correlation
between the frequency of $\sim 14$\,Myr old PMS stars and their distance 
from the massive $\sim 2$\,Myr old ionising objects in the field. This 
effect is not seen for non-PMS objects of similar brightness, confirming
that the trend is not due to problems in detecting faint objects
near the brightest stars. In Figure\,\ref{fig7} we show a map of the 
distribution of all PMS objects with respect to the 15 ionising stars. 
An inspection to Figure\,7 immediately
reveals two important facts: (i) there are very few PMS stars
near the barycentre of the massive objects and (ii) their $H\alpha$ 
luminosity is systematically lower than that of PMS stars farther
away. This means that the discs of the stars near the barycentre of the
ionising radiation are being eroded and the accretion process dwindles.
This is a clear example of how the younger generation of massive stars
can affect the late PMS evolution phases of somewhat older stars, right at 
the time when planets should start to form around them.  

\section{Conclusions}

In summary, even though our analysis is still in progress, we can
already draw some firm conclusions thanks to the very rich and 
homogeneous sample of PMS stars that we have put together.

It is clear that star formation is an ongoing process, since we see 
multiple generations of stars in all the fields that we have studied, 
separated by some 10\,Myr. Interestingly, the younger generations are
systematically more concentrated towards the center and the older 
generations are located farther out. No clear signs of ``triggering'' 
are present, although in most cases, the spatial density distribution 
of the older generations is consistent with an expansion velocity of some 
km\,s$^{-1}$ (a typical value in Galactic star forming regions) for a 
period of the order of 10-15\,Myr. Thus it is conceivable, although not
necessary, that also the older generations formed in the same place as
the more recent ones, some 10-15\,Myr ago.

We find that the mass accretion rate scales roughly with the square root
of the age and with the mass of the star to the power of $1.5$. The
physical conditions of the environment appear to have an effect on the
extent and duration of the star formation process, since the mass 
accretion rate for stars of the same mass and age is systematically higher 
in the Magellanic Clouds than in the Milky Way, and in the SMC than in the
LMC. The preliminary analysis that we have conducted so far to quantify 
this dependence indicates that metallicity has
an important effect on the rate of mass accretion and that the mass 
accretion rate scales roughly with the inverse of the cube root of the 
metallicity. One possible explanation for the role of metallicity in the
intensity and duration of the accretion process is that the lower 
radiation pressure exercised by the forming star on low-metallicity
disc material is less efficient at dispersing the disc and thus the
accretion process remains active for a longer time in lower-metallicity
environments. If this finding is confirmed when we extend this study to 
a wider range of environments, also including regions of less active star 
formation, it could have profound implications for the formation of
stars and planets in environments of low low-metallicity, such as those in
place in the early Universe.
 
\bigskip
\bigskip
\noindent {\bf Acknowledgments}\,\,\,
We are indebted to our collaborators Martino Romaniello, Giacomo Beccari, 
Loredana Spezzi, Elena Sabbi, Pier Prada Moroni, Scilla Degl'Innocenti, 
Francesco Paresce, and Morten Andersen for participating in this study
in the course of the years.

\bigskip
\bigskip
\noindent {\bf DISCUSSION}

\bigskip
\noindent {\bf BORIS SHUSTOV:} Evaporation is an important factor
limiting the final mass of stars not only at the stage of the circumstellar 
disc. It works earlier even in a more dramatic mode. When a dense
protostellar (starless) core is starting to be exposed to UV radiation
from nearby stars, or just to interstellar fields of UV photons, its
outer layer will evaporate, while the inner part will be stimulated to
form a young star by radiatively driven collapse. So the final mass of
the star or even its existence is the result of the competition between
these two processes.

\bigskip
\noindent {\bf GUIDO DE MARCHI:} Our observations at optical and near 
infrared wavelengths do not probe the earliest stages of star formation, 
since for those you would need the wavelength range offered by Herschel 
and particularly by Alma, which has the resolution needed to study in 
detail nearby star-forming regions. However, we clearly see the effects 
of the discs disruption due to photoevaporation at various stages of the 
PMS phase. Besides the example in the SN\,1987A field that I showed 
before (see Figure\,\ref{fig7}), we also see the effects of 
photoevaporation at play in the massive cluster 30\,Dor (as illustrated
in Figure 12 of De Marchi et al. 2011c). Also there we have a conspicuous 
population of older PMS stars, with ages of about 15 -- 20\, Myr, but the
situation is much more complex than in the field of SN\,1987A: not only are
there also other generations of PMS stars, but there is also a lot of gas in 
this region, much more than around SN\,1987A, some of which is cold molecular 
Hydrogen. What our observations show is that the stars with lowest 
$L(H\alpha)$ are preferentially in regions of lower gas density, while the 
stars with higher $L(H\alpha)$, are in denser regions. This suggests that the 
densest clouds are shielding these objects from the ionising radiation of the 
stars at the centre, while the discs of objects in less dense regions are 
at higher risk of evaporation. Thus, precisely as you said, the final mass 
of these stars is the result of the competition between many different 
processes that are at play simultaneously.

\end{document}